\documentclass[pre,twocolumn,aps,floatfix,longbibliography,natbib,nofootinbib]{revtex4-1}
\usepackage{pslatex,graphicx,dcolumn,bm,natbib,amssymb,amsmath,color,mathtools,mathrsfs}
\usepackage[utf8]{inputenc}
\usepackage{listings}
\usepackage[T1]{fontenc}

\newcommand{\be}{\begin{equation}}
\newcommand{\ee}{\end{equation}}

\begin{document}
\title{A biological tissue-inspired tunable photonic fluid}

\author{Xinzhi Li}
\author{Amit Das} 
\author{Dapeng Bi} 
\affiliation{Department of Physics, Northeastern University, MA 02115, USA}

\begin{abstract}
Inspired by how cells pack in dense biological tissues, we design 2D and 3D amorphous materials which possess a complete photonic bandgap. 
A physical parameter based on how cells adhere with one another and regulate their shapes can continuously tune the photonic bandgap size as well as the bulk mechanical properties of the material. The material can be tuned to go through a solid-fluid phase transition characterized by a vanishing shear modulus. Remarkably, the photonic bandgap persists in the fluid phase, giving rise to a photonic fluid that is robust to flow and rearrangements. Experimentally this design should lead to the engineering of self-assembled non-rigid photonic structures with photonic bandgaps that can be controlled in real time via mechanical and thermal tuning. 
\end{abstract}
\maketitle

Photonic bandgap (PBG) materials  have remained an intense focus of research since their introduction~\cite{John_PRL_1987,Yablonovitch_PRL_1987} and  have given rise to a wide range of applications such as  radiation sources~\cite{Cao_OPN_2005_review}, sensors, wave guides, solar arrays and optical computer chips~\cite{Chutinan_PRL_2003}. Most studies have been devoted to the design and optimization of photonic crystals -- a periodic arrangement of dielectric scattering materials that have photonic bands due to multiple Bragg scatterings. However, periodicity is not necessary to form PBGs and amorphous structures with PBGs~\cite{Noh_PRL_2011,Rechtsman_PRL_2011,Wiersma_NPHOTON_2013_review,Shi_AdvMater_2013_review} 
can offer many advantages over their crystalline counterparts~\cite{Wiersma_NPHOTON_2013_review}. 
For example amorphous photonic materials can exhibit bandgaps that are directionally isotropic~\cite{Florescu_PNAS_2009,Man_PNAS_2013} and are more robust to defects and errors in fabrication~\cite{Wiersma_NPHOTON_2013_review}. 
Currently there are few existing protocols for designing amorphous photonic materials. They include structures obtained from a dense packing of spheres (3D) or disks (2D)~
\cite{Yang_PRA_2010_ohern,Conley_PRL_2014,Muller_AOM_2014,Froufe_arxiv_2016_Scheffold_PRL,Noh_PRL_2011}, tailor-designed protocols that generate hyperuniform patterns~\cite{Florescu_PNAS_2009,Man_PNAS_2013,Froufe_arxiv_2016_Scheffold_PRL,Sellers2017} and spinodal-decomposed structures~\cite{Dufresne_SM_2009,Dong_PRE_2011}. While these designs yield PBGs, such structures are typically static, rigid constructions that do not allow tuning of photonic properties in real time and are unstable to structural changes such as large scale flows and positional rearrangements. 

In this work, we propose a new design for amorphous 2D and 3D PBG materials that is inspired by how cells pack in dense tissues in biology. We generate structures that exhibit broad PBGs based on a simple model that has been shown to describe cell shapes and tissue mechanical behavior. An advantage of this design is that the  photonic and mechanical properties of the material are closely coupled. The material can also be tuned to undergo a density-independent solid-fluid transition and the PBG persists well into the fluid phase. With recent advances in tunable self-assembly of nanoparticles or biomimetic emulsion droplets, this design can be used to create a `photonic fluid'.

\begin{figure*}[t]
\begin{center}
\includegraphics[width=2\columnwidth]{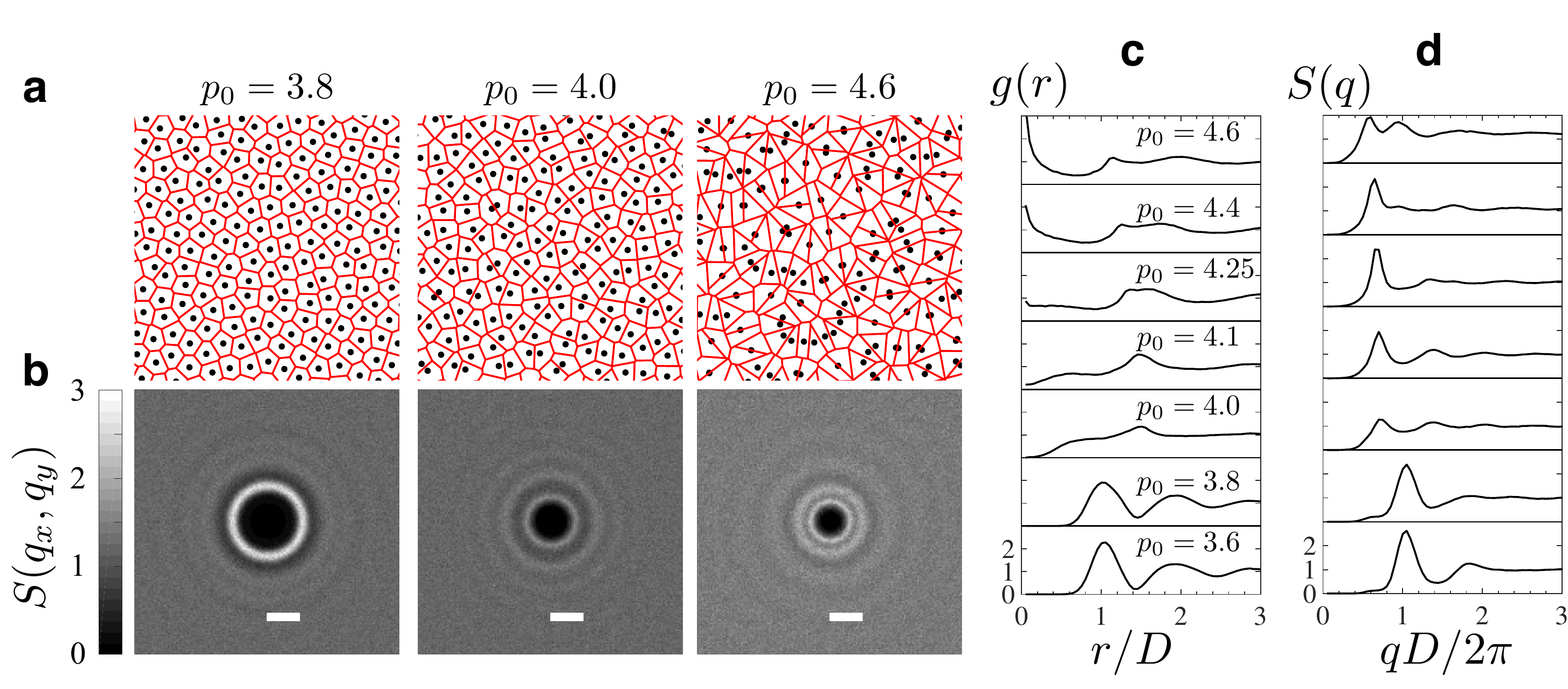}
\caption{
Tissue structure in the SPV model  . 
{\bf  \color{red} (a)}
 Simulation snapshots at 3 different values of the preferred cell perimeters $p_0$. Cell centers are indicated by points and cell shapes are given by their Voronoi tessellation (red outlines). 
{\bf  \color{red} (b)}
Contour plot of the structure factor $S(q_x,q_y)$ corresponding to the states shown in (a). Scale bar has length $2\pi/D$ in reciprocal space, where $D$ is the average spacing between cell centers.
{\bf  \color{red} (c)}
Pair-correlation function $g(r)$ at different values of $p_0$.
{\bf  \color{red} (d)}
Structure factor $S(q)$ at different values of $p_0$.
}
\label{fig:snapshot}
\end{center}
\end{figure*}

\subsection*{Model for epithelial cell packing in 2D}
When epithelial and endothelial cells pack densely in 2D to form a confluent monolayer, the structure of the resulting tissue can be described by a polygonal tiling~\cite{Ngain_PhilMagB_2001}. A great variety of cell shape structures have been observed in tissue monolayers, ranging from near-regular tiling of cells that resembles a dry foam or honeycomb lattice~\cite{Farhadifar_CurrBiol_2007} to highly irregular tilings of elongated cells~\cite{Duclos_SM_2014}. 
To better understand how cell shapes arise from cell-level interactions, a  framework called the Self-Propelled Voronoi (SPV) model has been developed recently~\cite{Bi_PRX_2016,cellgpu}.
In the SPV model, the basic degrees of freedom are the set of 2D cell centers $\{\bm r_i\}$ and  cell shapes are given by the resulting Voronoi tessellation. The complex biomechanics that govern intracellular and intercellular interactions can be coarse-grained~\cite{Ngain_PhilMagB_2001, Hufnagel_PNAS_2007, Farhadifar_CurrBiol_2007, Staple_EPJE_2010,Fletcher_BPJ_2014_vertex_review,Bi_SM_2014,Bi_NPHYS_2015} and expressed in terms of a mechanical energy functional for individual cell shapes. 
\be
E= \sum_{i=1}^N \left[ K_A (A_i-A_0)^2+ K_P (P_i-P_0)^2 \right].
\label{eq:total_energy}
\ee
The SPV energy functional is quadratic in both cell areas ($\{A_i\}$) with modulus $K_A$ and cell perimeters ($\{P_i\}$) with modulus $K_P$. The parameters $A_0$ and $P_0$ set the preferred values for area and perimeter, respectively. 
Changes to cell perimeters are directly related to the deformation of the acto-myosin cortex concentrated near the cell membrane. After expanding equation~\eqref{eq:total_energy},  the term $K_P P_i^2$ corresponds to the elastic energy associated with deforming the cortex. The linear term in cell perimeter, $- 2 K_P P_0 P_i$, represents the effective line tension in the cortex and gives rise to a `preferred perimeter'  $P_0$. The value of $P_0$ can be decreased by up-regulating  the contractile tension in the cortex~\cite{Farhadifar_CurrBiol_2007, Staple_EPJE_2010,Bi_NPHYS_2015} and it can be increased by up-regulating cell-cell adhesion.
We simulate tissues containing $N$ cells under periodic boundary conditions, the value of $N$ has been varied from $N=64$ to $1600$  to check for finite-size effects (SI Appendix Fig.~S6) . $A_0$ is set to be equal to the average area per cell and $\sqrt{A_0}$ is used as the unit of length. 
After non-dimensionalizing Eq.~\ref{eq:total_energy} by  $K_A A_0^2$ as the unit energy scale, we  choose $K_P /(K_A A_0)= 1$ such that the perimeter and area terms contribute equally to the cell shapes. The choice of $K_P$ does not affect the results presented.   The preferred cell perimeter is rescaled $p_0=P_0/\sqrt{A_0}$ and varied between $3.7$ (corresponding to the perimeter of a regular hexagon with unit area) and $4.6$ (corresponding to the perimeter of an equilateral triangle with unit area) ~\cite{Bi_NPHYS_2015}. We obtain disordered ground states of the SPV model by minimizing $E$ using the L-BFGS method~\cite{L-BFGS} starting from a random Poisson point pattern. 
We also test the finite temperature behavior of the SPV model by performing Brownian dynamics~\cite{Bi_PRX_2016}.  

The ground states of Eq.~\ref{eq:total_energy} are amorphous tilings where the cells have approximately equal area, but varying perimeters as dictated by the preferred cell perimeter $p_0$. It has been shown that at a critical value of the   $p_0^* \approx 3.81$, the tissue collectively undergoes a solid-fluid transition~\cite{Bi_NPHYS_2015}. 
When $p_0<p_0^*$, cells must overcome finite energy barriers to rearrange and the tissue behaves as a solid , while above $p_0^*$, the tissue becomes a fluid with a vanishing shear modulus as well as vanishing energy barriers for rearrangements~\cite{Bi_NPHYS_2015}.
Coupled to these mechanical changes, there is a clear  signature in cell shapes at the transition~\cite{Park_NMAT_2015,Bi_PRX_2016,Atia2018}: the shape-based order parameter calculated by averaging the observed cell perimeter-to-area ratio $s = \langle P/\sqrt{A} \rangle$ grows linearly with $p_0$ when  $p_0>p_0^*$ in the fluid phase but remains at a constant ($s\approx p_0^*$) in the solid phase.  In Fig.~\ref{fig:snapshot}(a) we show three representative snapshots of the ground states at various values of $p_0$.  We take advantage of the diversity and tunability of the point patterns and cell structures (Fig.~\ref{fig:snapshot}(a)) produced by the SPV model and use them as templates to engineer photonic materials.

\subsection*{Characterization of 2D structure}
To better understand the ground states of the SPV model,
we first probe short-range order by analyzing the pair-correlation function $g(r)$ (see details in Materials and Methods)  of  cell centers . In Fig.~\ref{fig:snapshot}(c), $g(r)$ for the solid phase ($p_0<3.81$) shows mutual exclusion between nearest neighbors and becomes constant at large distances. These features are similar to those observed in other amorphous materials with short-range repulsion and a lack of long range positional order, such as jammed granular packings ~\cite{Ohern_PRE_2003} and dense colloidal arrays~\cite{Kegel_Science_2000}. 
When $p_0$ is increased, the tissue enters into the fluid phase at $p_0 > 3.81$. In order to satisfy the higher preferred perimeters, cells must become more elongated. And when two neighboring elongated cells align locally, their centers can be near each other whereas cells not aligned will have centers that are further apart. As a result, the first peak of $g(r)$ broadens. Hence as $p_0$ is increased, short range order is reduced. 
Deeper in the fluid regime, when $p_0 > 4.2$, $g(r)$ starts to develop a peak close to $r=0$ which means that cell centers can come arbitrarily close. 
Interestingly the loss of short range order does not coincide with the solid-fluid transition, which yields  an intermediate fluid state that retains short range order.

Next we focus on the structural order at long lengthscales. While the SPV ground states are aperiodic by construction, they show interesting long range {\emph{density}} correlations.  
 The structure factor $S({\bf{q}})$ (see details in Materials and Methods) is plotted for various $p_0$ values in Figs.~\ref{fig:snapshot}(b) \& (d). Strikingly, for all  $p_0$ values tested, the structure factor vanishes as ${\bf{q}} \to 0$, corresponding  to a persistent density correlation at long distances.  This type of `hidden' long range order is characteristic of patterns that are known as   hyperuniformity~\cite{Torquato_Stillinger_PRE_2003_Hyperuniformity,Zachary_JSTAT_2009_HU}. 
In real space,  hyperuniformity is equivalent to when the variance of the number of points $\sigma_R^2$ in an observation window of radius $R$ grows as function of the surface area of the window, i.e.  $\sigma_R^2  \propto R^{d-1}$, 
where $d$ is the space dimension~\cite{Torquato_Stillinger_PRE_2003_Hyperuniformity}. 
This is in contrast to the  $\sigma_R^2 \propto R^{d}$ scaling that  holds for  uncorrelated random patterns. 
Indeed, real space measurements in the SPV model also confirm that the distribution of cell centers is strongly hyperuniform for all values of $p_0$ and system sizes tested that include both solid and fluid states  (SI Appendix Fig.~S1).

It has been suggested that  hyperuniform amorphous patterns can be used to design photonic materials that yield PBGs ~\cite{Florescu_PNAS_2009,Man_PNAS_2013}. Florescu \& coworkers~\cite{Florescu_PNAS_2009} further conjectured that hyperuniformity is \emph{necessary} for the creation of PBGs. However, recent work by  Froufe-P{\'e}rez et al~\cite{Froufe_arxiv_2016_Scheffold_PRL}  have demonstrated that short-range order rather than hyperuniformity may be more important for PBGs. The SPV model provides a unique example of a hyperuniform point pattern with a short-range order that can be turned on-and-off.  This tunability will allow a direct test of the ideas proposed in refs~\cite{Florescu_PNAS_2009,Man_PNAS_2013,Froufe_arxiv_2016_Scheffold_PRL}.

\begin{figure}[htbp]
\begin{center}
\includegraphics[width=1\columnwidth]{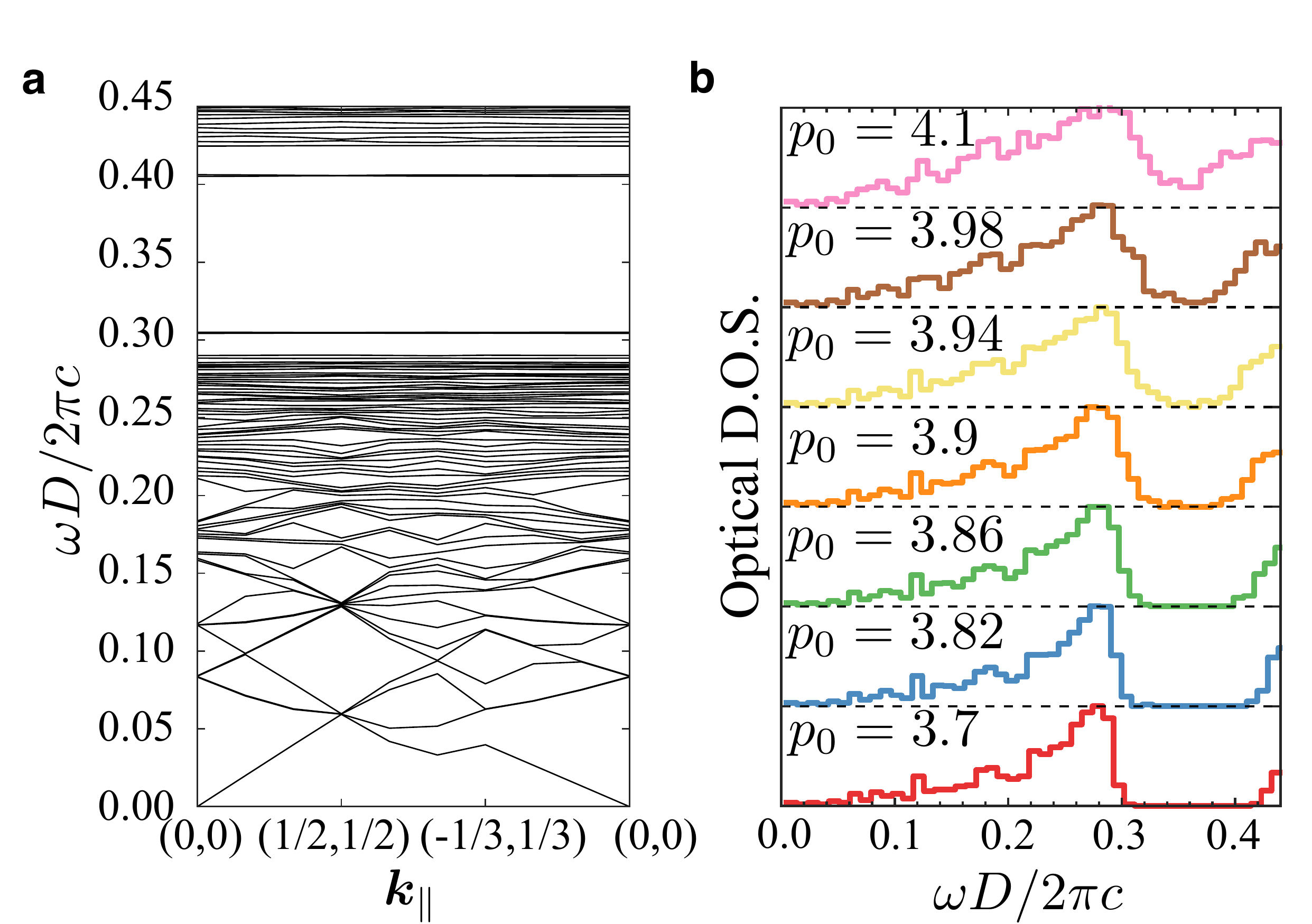}
\caption{
Characterizing photonic properties in 2D.
{\bf  \color{red} (a)}
Photonic band structure of Transverse Magnetic(TM) for the material constructed by placing dielectric cylinders at the cell centers exhibits a PBG. Design is based on a ground state of the SPV at $p_0=3.85$. $\bm{k}_\|$ is the in-plane wave vector.   
{\bf  \color{red} (b)}
The  optical density of states of TM at different values of $p_0$. The width of the bandgap $\Delta \omega$ has a strong dependence on $p_0$, while the midgap frequency $\omega_0$ remains constant  (SI Appendix Fig. S4). 
}
\label{fig:tm}
\end{center}
\end{figure}

\subsection*{2D Photonic Material Design and Properties}
For any point pattern (crystalline or amorphous), the first step in the engineering of PBGs is to decorate it with a high dielectric contrast material. 
 The simplest protocol is to place cylinders centered at  each point $\{ \bm{r}_i=(x_i,y_i) \}$  that are infinitely tall in the $z-$direction. Such design typically  yields  bandgaps in the Transverse Magnetic (TM) polarization (the magnetic field is parallel to the xy-plane)~\cite{Joannopoulos_book_2011}. 
Based on this design, we first construct a material using SPV point patterns. In order to maximize the size of the bandgap, the cylinders are endowed with dielectric constant $\epsilon=11.56$ and radius $r/D=0.189$~\cite{Froufe_arxiv_2016_Scheffold_PRL}. We will refer to this construction as ``TM-optimized". 
We also use a second decoration method~\cite{Florescu_PNAS_2009,Froufe_arxiv_2016_Scheffold_PRL}  which has been shown to yield complete  PBGs, i.e. gaps in both TM and Transverse Electric (TE) polarizations. We use a design based on the Delaunay triangulation of a point pattern. Cylinders with $\epsilon=11.56$ and radius $r/D=0.18$ are placed at the nodes of the Delaunay triangulation while walls with  $\epsilon=11.56$ and thickness $w/D=0.05$ are placed on the bonds of this trivalent network.  We refer to this construction as ``TM+TE optimized".
Photonic properties are numerically calculated using the plane wave expansion method~\cite{Johnson_OptExp_2001} implemented in the MIT Photonic Bands program. We use the supercell approximation in which a finite sample of $N$ cells is repeated periodically. 
The photonic band structure is calculated by following the path of  
${\bm{k}_\|} = (0,0) \to  (\frac{1}{2},\frac{1}{2}) \to (-\frac{1}{3},\frac{1}{3}) \to (0,0) $
in reciprocal space.  The  SI Appendix includes a sample script used for photonic band calculations.

\begin{figure*}[t]
\begin{center}
\includegraphics[width=2\columnwidth]{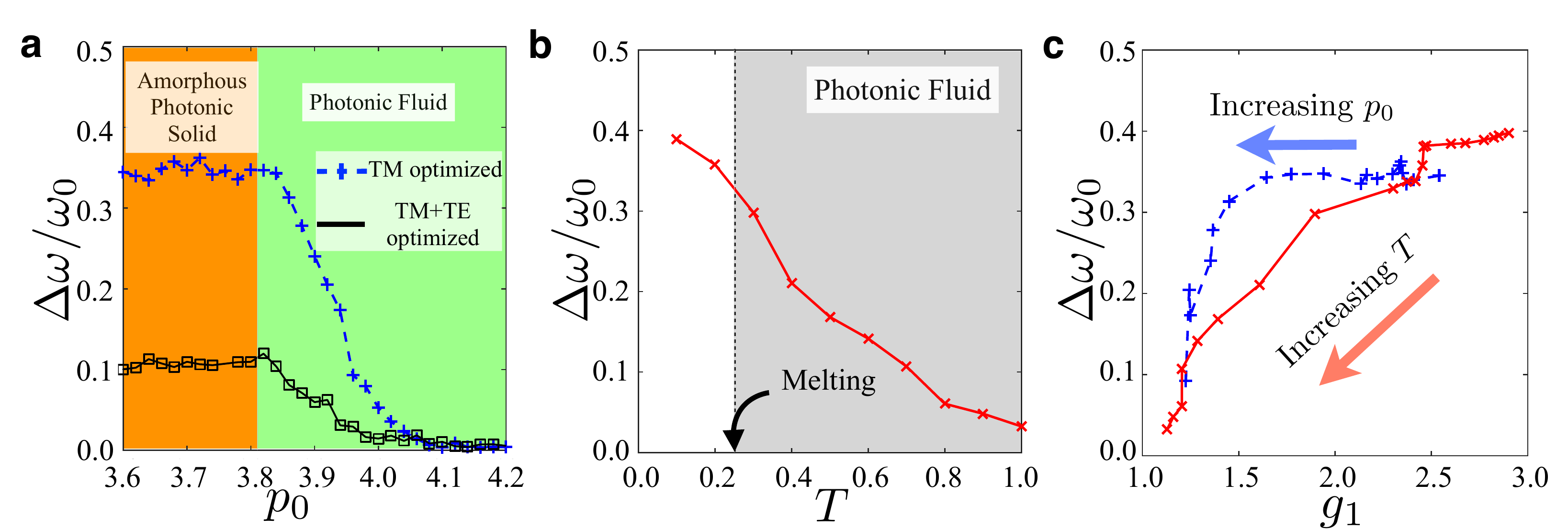}
\caption{
{\bf  \color{red} (a)}
 The structure of the PBG as function of $p_0$. 
 The ''TM-optimized'' structure corresponds to the size of the TM bandgap for a material constructed by placing  dielectric cylinders at cell centers. 
 The ''TM+TE Optimized'' structure corresponds to complete PBGs for a material constructed using a trivalent network design (see text). 
Phases are colored according to the mechanical property of the material. At $p_0 \approx 3.81$ the material under goes a solid-fluid transition where the shear modulus vanishes. The dependence of $\omega_0$ on $p_0$ is weak as shown in  SI Appendix  Fig.~S4(a).
{\bf  \color{red} (b)}
Effect of heating on the PBG in the TM optimized structure. At fixed $p_0=3.7$, temperature T is gradually increased. At $T \approx 0.25$ the material begins to fluidize through melting while the bandgap still persists into the mechanical fluid phase.   
{\bf  \color{red} (c)} 
The relationship  between the bandgap size in the TM optimized structure and the short range order of the system. The explicit dependence of short range order on $p_0$ and $T$ is shown in  SI Appendix Fig.~S2 and S3.
}
\label{fig:pd}
\end{center}
\end{figure*}

The TM-optimized band structure based on a SPV ground state with $N=64$ cells and $p_0 = 3.85$ is shown in Fig.~\ref{fig:tm}(a). 
Due to the aperiodic nature of the structure, the PBG is isotropic in $\bm{k}_\|$.   We also calculate the optical density of states (ODOS) (Fig.~\ref{fig:tm}(b)) by binning eigenfrequencies from $10$ samples with the same $p_0$ but different initial seeds.  
The relative size of the PBG can be characterized by the  gap-midgap ratio $\Delta \omega / \omega_0$, which is plotted as function of $p_0$ in Fig.~\ref{fig:pd}(a) for both TM and TM+TE optimized structures. We find that the size of the PBG is constant in the solid phase of the SPV model ($p_0<p_0^*$) with width $\Delta \omega / \omega_0 \approx 0.36$ for the TM-optimized structure and $\Delta \omega / \omega_0 \approx 0.1$ in the TM+TE optimized structure. In the fluid phase, $\Delta \omega / \omega_0$ decreases as $p_0$ increases, yet stays  finite in the range of $3.81<p_0 \lesssim 4.0$. At even larger $p_0$ values, the PBG vanishes. The location at which the PBG vanishes appears to coincide with the loss of short-range order in the structure. To quantify this, we plot $\Delta \omega / \omega_0$  vs the first peak height of the pair-correlation $g_1$ in Fig.~\ref{fig:pd}(c). This is in agreement with the findings of Yang et al~\cite{Yang_PRA_2010_ohern}, and suggests that  short-range positional order is essential to obtaining a PBGs which allows for collective Bragg backscattering of the dielectric material. \emph{This also shows that PBGs are absent in states which are hyperuniform but missing short-range order.}  

To test for finite-size dependence of the ODOS and the PBG,
we carry out the photonic band calculations at $p_0=3.7$ for various system sizes ranging from $N=64$ to $N=1600$.
At each system size, the ODOS was generated by tabulating TM frequencies along the  ${\bm{k}_\|} = (0,0)$ and ${\bm{k}_\|} = (0.5,0.5)$ directions in reciprocal space for $10$ randomly generated states. As shown in  SI Appendix Fig.~S6 , while low frequency modes may depend on the system size that bigger fluctuations exist for smaller systems, the PBG is always located between mode number $N$ and $N+1$ and has a width that does not change as function of $N$.

 Since the model behaves as a fluid-like state above $p_0^*\approx3.81$~\cite{Bi_NPHYS_2015,Bi_PRX_2016} and the PBG does not vanish in the fluid-like regime Fig.~\ref{fig:pd}(a), 
this gives rise to a {\bf photonic fluid} where a PBG can exist without a static and rigid structure. To show this explicitly, we test the effects of fluid-like cell rearrangements by analyzing the photonic properties of the   dynamical fluid phase at finite temperature.

At a fixed value of $p_0=3.7$, we simulate Brownian dynamics in the SPV model at different temperatures $T$. At each $T$, we take $10$ steady state samples and construct the TM-optimized structures to calculate their ODOS. In Fig.~\ref{fig:pd}(b), we plot the bandgap size as function of increasing temperature. Note that even past the melting temperature of $T\approx0.25$~\cite{Bi_PRX_2016}, the PBG does not vanish, again giving rise to a robust photonic fluid phase. Inside the fluid phase, we also find that the PBG is not affected by the positional changes due to cell rearrangements  (SI Appendix Fig. S5). Finally, we analyze the effect of heating on the short-range order. In  Fig.~\ref{fig:pd}(c), increasing $T$ results in another `path' in manipulating the short-range order.

\subsection*{Extension to 3D photonic material}

\begin{figure*}[t]
\begin{center}
\includegraphics[width=2\columnwidth]{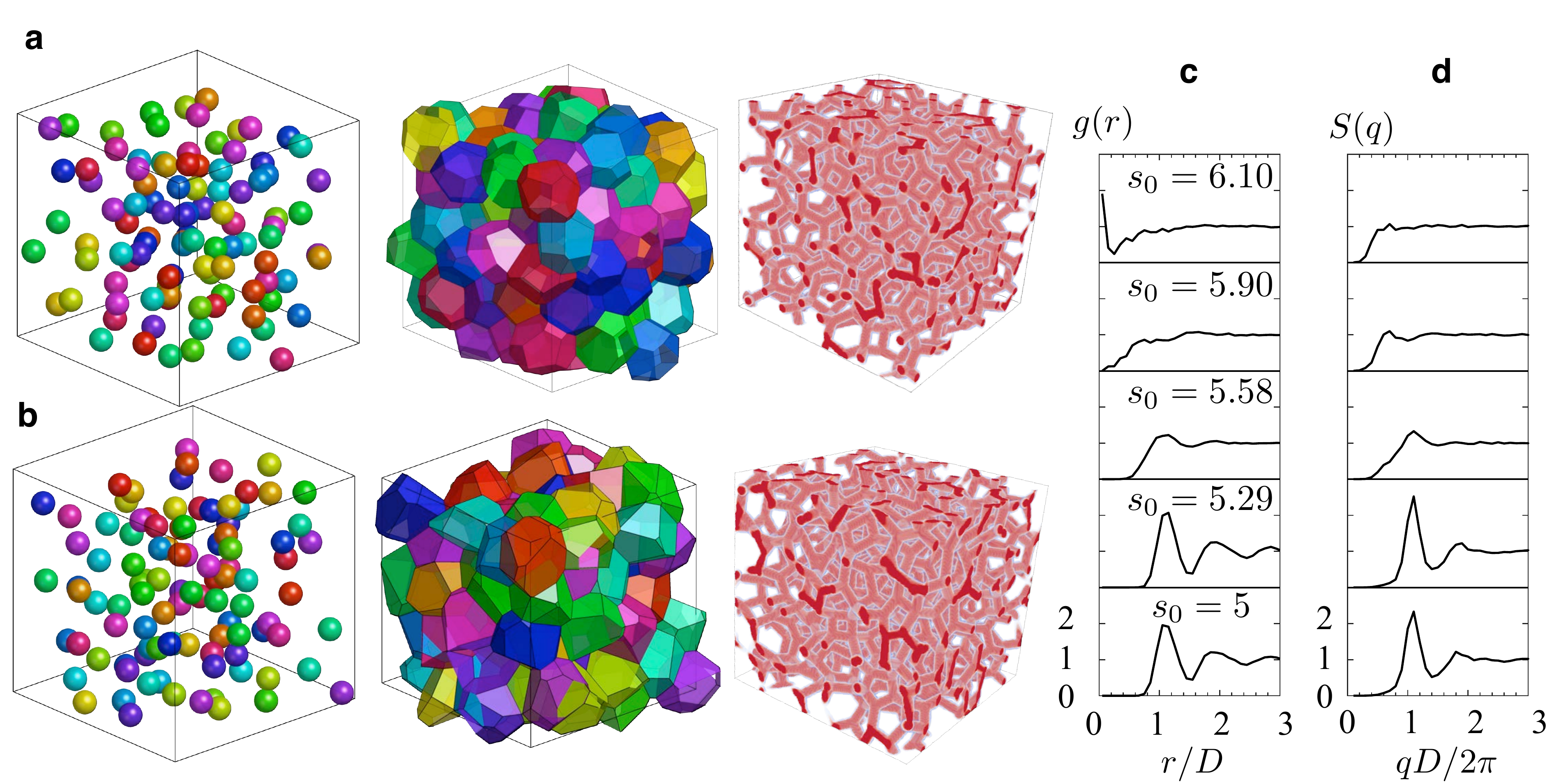}
\caption{
Structure characterization in the 3D Voronoi-cell model.
{\bf  \color{red} (a)} Cell centers and their corresponding Voronoi tessellation as well as the decorated 3D photonic structure
are shown for a state at $s_0=5$ and $N=100$ cells. The cell centers are drawn with a finite size and different colors only to aid  visualization. 
{\bf  \color{red} (b)} Cell centers and their corresponding Voronoi tessellation as well as the decorated 3D photonic structure
are shown for a state at $s_0=5.82$ and $N=100$ cells.
{\bf  \color{red} (c)}
Pair-correlation function $g(r)$ at different values of $s_0$.
{\bf  \color{red} (d)}
Structure factor $S(q)$ at different values of $s_0$.
}
\label{fig:3d_schematic}
\end{center}
\end{figure*}

In order to further demonstrate the viability and versatility of tissue-inspired structures as design templates for photonic materials, we extend this study to 3D.  Recently,  Merkel and Manning~\cite{Merkel2017,sharp_2018}  generalized the 2D SPV model to simulate cell shapes in 3D tissue aggregates by replacing the cell area and perimeter with the cell volume and surface area, respectively. 
This results in a quadratic energy functional that is a direct analog of Eq. ~\ref{eq:total_energy}
\be
E= \sum_{i=1}^N \left[ K_S(S_i-S_0)^2 +  K_V(V_i-V_0)^2 \right],
\label{eq:total_energy_3D}
\ee
where $S_i$ and $V_i$ are the surface area and volume of the \textit{i}-th cell in 3D,  with the preferred surface area and volume being $S_0$ and $V_0$, respectively. Similar to the 2D version of the model, we have two moduli - $K_S$ for the surface area and $K_V$ for volume. Following~\cite{Merkel2017}, we make our model dimensionless by setting $V_{0}^{1/3}$ as the unit of length and $K_{S}V_{0}^{4/3}$ as the unit of energy. This gives us a  dimensionless energy and dimensionless shape factor $s_{0}=S_{0}/V_{0}^{2/3}$ as the single tunable parameter in 3D which again is the anaolog of the parameter $p_0$ is the 2D scenario. For the 3D model, we perform Monte-Carlo simulations~\cite{landau2014guide} with the cell centers at a scaled temperature $T=1$, expressed in the unit $K_{S}V_{0}^{4/3}/k_{B}$ where $k_B$ is the Boltzmann constant. During the simulation, each randomly chosen cell center is given by a displacement  following the Metropolis algorithm~\cite{Allen:1989} where the Boltzmann factor is calculated using the energy function in Eq.~\ref{eq:total_energy_3D}. For this purpose, we extract the cell surface areas and volumes from 3D Voronoi tessellations generated using the Voro++ library~\cite{voro_pp}. $N$ such moves constitute a Monte-Carlo step and we perform $10^{5}$ such steps to ensure that the cells have reached a steady state. Then we run for another $10^5$ steps to compute the $g(r)$ and $S(q)$.
The average volume per cell is held constant for this procedure. Periodic boundary conditions were applied in all directions.  For all the simulations we keep the scaled modulus $K_{V}V_{0}^{2/3}/K_{S} = 1$.
It was previously found~\cite{Merkel2017} that the $T=0$ ground states undergo a similar solid-to-fluid transition at $s_0 \sim 5.4$, i.e. solid for $s_0<5.4$ and fluid when $s_0\ge5.4$.  Our Monte-Carlo simulations also recover this transition near $s_0=5.4$ when we plot an effective diffusivity, extracted from the mean-aquare displacement(MSD) of the cells, at different $s_0$ values  (SI Appendix Fig.~S9).  States below the transition  ($s_0=5$)  and above the transition  ($s_0=5.82$) are shown in Fig.~\ref{fig:3d_schematic} for a system of $N=100$ cells.

We characterize the structure of the 3D cell model by calculating their pair-correlation function $g(r)$. In Fig.~\ref{fig:3d_schematic}(c), the short range order behaves similar to the 2D case. At low values of $s_0$ that correspond to rigid solids, the first peak in $g(r)$ is pronounced suggesting an effective repulsion between nearest neighbors. As $s_0$ is increased, this short-range order gets eroded as a preference for larger surface area to volume ratio allows  nearest neighbors to be close. The $s_0=6.1$ state in Fig.~\ref{fig:3d_schematic}(c) is an extreme example where it is possible for two cell centers to be arbitrarily near. However the long-range order remains  throughout all values of $s_0$ tested as shown by the $S(q)$ in Fig.~\ref{fig:3d_schematic}(d). The small value of $S(q\to0)$ is a consequence of suppressed density fluctuations across the system when cells all prefer the same volume dictated by Eq.~\ref{eq:total_energy_3D}. Whether they are truly hyperuniform would require sampling at higher system sizes, which is beyond the scope of this study.

\begin{figure}[ht]
\begin{center}
\includegraphics[width=1\columnwidth]{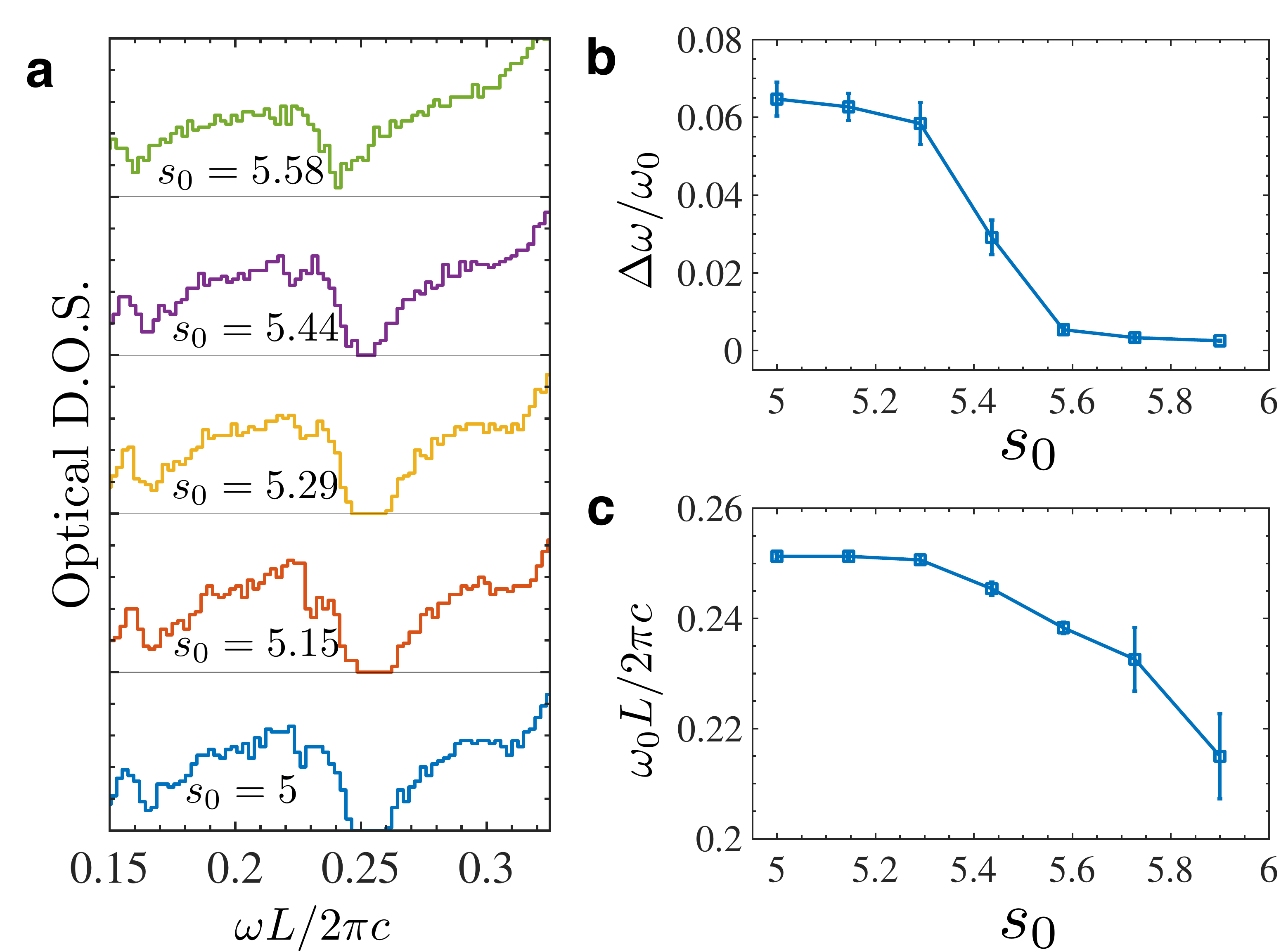}
\caption{
Characterizing photonic properties in the 3D design. 
{\bf  \color{red} (a)}
The optical density of states is shown at various values of $s_0$. Here frequency is in units of $2\pi c/ L$, where $L$ is the average edge length in the photonic network.  
{\bf  \color{red} (b)}
The size of the gap-midgap ratio $\Delta \omega / \omega_0$ as function of $s_0$. 
{\bf  \color{red} (c)}
Midgap frequency $\omega_0$ as function of $s_0$. 
}
\label{fig:3d_results}
\end{center}
\end{figure}

Next, in order to make a photonic material we decorate the Voronoi tessellations to create a connected dielectric network. While a Voronoi tessellation is already a connected network made of vertices and edges, it possess a large dispersion of edge lengths. This can result in vertices that are arbitrarily close to each other and could  hinder the creation of PBGs~\cite{Liew_PRA_2011}. To overcome this we adopt a method described in ~\cite{Sigmund_Hougaard_PRL_2008,Florescu_PNAS_2009, Sellers2017} to make the structure locally more uniform. 
In a 3D Voronoi tessellation, each vertex is calculated from the circumcenter of the 4 neighboring cell centers and edges are formed by connecting adjacent vertices. 
In this design protocol, the connectivity of the dielectric network is the same as the network of the vertices and edges in the Voronoi tessellation. However, the  vertex positions of the dielectric network are replaced by the center-of-mass (barycenters) of the 4 neighboring Voronoi cell-centers. The resulting structure is  a tetrahedrally connected  network where the edges are more uniform in length, two representative samples are shown in Fig.~\ref{fig:3d_results}(a). 
Next we decorate the network with dielectric rods of width $W$ running along the edges.  For the dielectric rods we again use $\epsilon=11.56$ and their width $W$ is chosen such that the volume filling fraction of the network is $V_\text{rod} / V_\text{box} = 20\%$~\cite{Liew_PRA_2011}.

 The photonic properties of the 3D dielectric network are calculated using the MIT Photonic Bands program~\cite{Johnson_OptExp_2001}. We calculate the ODOS for structures based on different values of $s_0$. Here we have chosen structures containing $N=100$ cells. Due to the isotropic nature of the photonic band structure  (SI Appendix Fig.~S7) , we generate the ODOS based on two reciprocal vectors ${\bf{k}} = (0,0,0) \ \& \ {\bf{k}} =(0.5,0.5,0)$. At each value of $s_0$, we average over 10 different random samples. We find the first complete photonic bandgap between mode number $n_V$ and $n_V+1$, where $n_V$ is the number of vertices. In the solid phase of the model ($s_0<5.4$), we find an average  gap-midgap ratio of  $\sim 6\%$, this decreases as $s_0$ is increased and becomes vanishingly small when the model is in its fluid phase ($s_0>5.4$) (Fig.~\ref{fig:3d_results}(b)). Interestingly, the midgap frequency $\omega_0$ also shifts slightly toward lower values with increasing $s_0$(Fig.~\ref{fig:3d_results}(c)). The appearance of the PBG here also coincides with the presence of short-range order. The PBG vanishes when there is no longer a pronounced first peak in the $g(r)$.

\section*{Discussion}
We have shown that structures inspired by how cells pack in dense tissues  can be used as a template for designing amorphous materials with full photonic bandgaps. The most striking feature of this material is the simultaneous tunability of mechanical and photonic properties. The structures have a short-range order that can be tuned by a single parameter, which governs the ratio between cell surface area and cell volume (or perimeter-to-area ratio in 2D). The resulting material can be tuned to transition between a solid and a fluid state and the PBG can be varied continuously. Remarkably, the PBG persists even when the material behaves as a fluid. 
Furthermore, we have explored different ways of tuning the short-range order in the material including cooling/heating and changing cell-cell interactions. We propose that the results in Fig.~\ref{fig:pd}(c) can be used as a guide map for building a photonic switch that is either mechano-sensitive (changing $p_0$ or $s_0$) or thermosensitive (cooling/heating).

While they are seemingly devoid of long-range order (i.e. they are always amorphous and non-periodic), these tissue inspired structures always exhibit strong hyperuniformity. Most interestingly, there are two classes of hyperuniform states found in this work: one that has short-range order and one that does not. While the former is similar to hyperuniform structures studied previously, the latter class is new and exotic which has not been observed before. Furthermore, we have shown that hyperuniformity alone is not sufficient for obtaining PBGs, which complements recent studies~\cite{Froufe_arxiv_2016_Scheffold_PRL}. Rather, the presence of short-range order is crucial for a PBG. 

Another recent study has suggested that the local self-uniformity~\cite{Sellers2017} (LSU) -- a measure of the similarity in a network's internal structure is crucial for bandgap formation. We believe that the states from the SPV model with short-range order also have a high degree of LSU and it is likely that LSU is a more stringent criterion for PBG formation than the simplified measure of short-range order based on $g(r)$. This will be a promising avenue for future work. 

It will be straightforward to manufacture static photonic materials based on this design using 3D printing or laser etching techniques~\cite{Man_PNAS_2013,Muller_AOM_2014,Sellers2017}.  In addition, top-down fabrication techniques such as electron beam lithography or focused ion beam milling can also be used which will precisely control the geometry and arrangement of micro/nanostructures.

An even more exciting possibility is to adapt this design protocol to self-assemble structures. 
 Recent advances in emulsion droplets have demonstrated feasibility to reconfigure the droplet network via tunable interfacial tensions and bulk mechanical compression~\cite{Pontani_PNAS_2012, Zarzar2015, Pontani_biophysj_2016}. 
 Tuning interfacial interactions such as tension and adhesion coincides closely with the tuning parameter in the cell-based model $p_0$ (2D) and $s_0$ (3D) which are inspired by the interplay of intracellular tension and cell-cell adhesion in a biological tissue.  
 Experimentally this may lead to construction of  a tunable range of artificial cellular materials that vary in  surface area-to-volume ratios. Further the dynamically reconfigurable complex emulsions via tunable interfacial tensions~\cite{Zarzar2015} could lead to switching of  these properties in real time. Another possibility is the creation of inverted-structures or ``inverse opals" by using the emulsion droplet network as a template~\cite{Imhof_Pine_emulsion_template,emulsion_templating_review}, which is often able to enhance the photonic properties over the original template.  Alternatively, nanoparticles grafted with polymer brushes~\cite{Kumar_Macromol_2013,Rogers_NatRevMat_2016} could be desgined to tune the effective surface tension of individual particles as well as the adhesion strengths between particles. This could allow the mimicking of the interaction between cells and give rise to a controllable preferred cell perimeter in 2D. 

 Two recent open-source software packages:  cellgpu~\cite{cellgpu} and  AVM~\cite{avm_model} have made it convenient to simulate very large system sizes in 2D. This would allow in-depth analysis of the system-size dependence of long and short-range order in this model, which is an exciting possibility for future research.

\section*{Material and Methods}
\subsection*{Characterization of structure}
The pair correlation function is defined as~\cite{Theory_Simple_Liquids_gr}$$g(r)=\frac{1}{N\rho}\left \langle \sum\limits_{i} \sum\limits_{i \neq j} \delta(\vec{r}+\vec{r_i}-\vec{r_j })\right \rangle$$ $\vec{r_i}$ is the position of the $i^{\text{th}}$ particle, $N$ is number of particle and $\rho$ is the average particle density of the isotropic system. $\left \langle \cdot \right \rangle$ denotes $n$ ensemble average over different random configurations at the same parameter set. 
The  structure factor is given by~\cite{Theory_Simple_Liquids_Sq}
$$S(\vec{q}) = \frac{1}{N}\left \langle\sum\limits_{i, j} e^{i\vec{q}\cdot(\vec{r_i}-\vec{r_j })} \right \rangle$$
$\left \langle \cdot \right \rangle$ denotes $n$ ensemble average over different random configurations at the same parameter set.
$S(q)$ is obtained by further averaging over a sphere of radius $q$ in reciprocal space. 

\subsection*{Photonic band structure calculation}
Photonic properties are numerically calculated using the plane wave expansion method~\cite{Johnson_OptExp_2001} implemented in the MIT Photonic Bands program. We use the supercell approximation in which a finite sample of $N$ cells is repeated periodically. The  SI Appendix  includes a sample script used for photonic band calculations in 2D.

\section*{Acknowledgements}
The authors wish to thank M. Lisa Manning, Bulbul Chakraborty, Abe Clark, Daniel Sussman and Ran Ni for helpful discussions.
The authors acknowledge the support of the Northeastern University Discovery Cluster.

\bibliography{photonic}

\end{document}